\documentclass[12pt,a4paper,prc,showpacs,showkeys,preprintnumbers,unsortedaddress,amsmath,amssymb,floatfix]{revtex4}

\input epsf.sty
\usepackage{graphicx}
\usepackage{bm}
\usepackage{graphicx}

\begin{document}

\title{Soft Dipole Modes in Neutron-rich Ni-isotopes in QRRPA}

\author{Li-Gang Cao$^{1}$ and Zhong-Yu Ma$^{2,3}$\footnote{also Center of Theoretical
Nuclear Physics, National Laboratory of Heavy Ion Accelerator of
Lanzhou, Lanzhou 730000 and Institute of Theoretical Physics,
Chinese Academy of Sciences, Beijing 100080}}

\affiliation{$^{1}$Institute of High Energy Physics, Chinese
Academy of Sciences, Beijing 100039, P.R. of China}

\affiliation{$^{2}$ China Center of Advanced Science and
Technology (World Laboratory), P.O.Box 8730, Beijing 100080, P.R.
of China}

\affiliation{$^{3}$China Institute of Atomic Energy, Beijing
102413, P.R. of China}

\date{\today}

\begin{abstract}

The soft dipole modes in neutron rich even-even Ni-isotopes are
investigated in the quasiparticle relativistic random phase
approximation. We study the evolution of strengths distribution,
centroid energies of dipole excitation in low-lying and normal GDR
regions with the increase of the neutron excess. It is found in
the present study that the centroid energies of the soft dipole
strengths strongly depend on the thickness of neutron skin along
with the neutron rich even-even Ni-isotopes.

\end{abstract}



\maketitle

\section{introduction}

Currently there is, in nuclear physics community, a strong
interest in the study of weakly bound nuclei both experimentally
and theoretically. The experiments with the radioactive ion beam
provide the possibility to study the properties of weakly bound
nuclei with a large neutron or proton deficient. One of the most
hot issues on the weakly bound nuclei, presently, is focused on
the study of multipole response of nuclei far from
$\beta$-stability line, especially in the isovector dipole mode. A
genuine feature of weakly bound nuclei is the appearance of soft
electric dipole modes, so-called Pygmy Dipole Resonances. The
appearances of low-lying dipole modes in weakly bound nuclei are
of particular interest because they are expected to reflect the
motion of neutron skin against the core formed with equal number
of neutrons and protons as well as to provide important
information on the isospin and density-dependent parts of the
effective interactions. Over the last decade, much effort has been
dedicated to the investigation of the properties of soft dipole
modes in light neutron rich
nuclei\cite{Try03,Rye02,Lei01,Iwa00,Iwa001,Aum99,Sac93,Nak94,Aum96,Vre01,Sag01}.
In order to obtain more valuable information for such a peculiar
mode, detail studies of the low-lying dipole excitation over
sufficiently long isotopic chains are required.

In neutron rich nuclei, the weakly bound outermost neutrons form
neutron skin at the nuclear surface. The investigation of neutron
skin has been paid more attentions due to the fact that the
thickness of neutron skin are related to the equation of state
(EoS) of asymmetric nuclear matter near saturation density, which
is an essential ingredient in the calculation of the properties of
neutron stars\cite{Bro00,Furn02,Bod03,Yos04}. Various experiments
have been performed or proposed to measure the neutron density
distribution and therefore, the differences between radii of the
neutron and proton
distribution\cite{Die03,Tso03,Vre03,Hor01,Bat89,Suz95,Kra91,Kra99}.
Although the neutron density distribution is difficult to be
measured, the experiments on the giant dipole resonance (GDR) or
spin-dipole resonance (SDR) excitation may provide some
information of the difference between neutron and proton
radii\cite{Kra91,Kra99}.

In order to describe the collective excitations in exotic nuclei,
a number of theoretical works have been devoted to study the
properties of multipole responses in open shell nuclei by taking
into account the effect of pairing correlations in the framework
of quasiparticle random phase approximation
(QRPA)\cite{Kam98,Kha00,Mat01,Hag01,Pa03}. In this paper, we aim
at the investigation of the evolution of low-lying dipole modes of
neutron rich even-even Ni-isotopes. The method we used is the
quasiparticle relativistic random phase approximation (QRRPA),
which has been formulated in the response function
method\cite{Cao032}. We first investigate the ground state
properties of neutron rich Ni-isotopes in the extended RMF+BCS
approximation\cite{Cao031}, where the contribution of the resonant
continuum to the pairing correlations is  treated properly. Based
on the ground states in RMF+BCS calculations, we construct the
configurations of quasiparticle excitations, solve the
Bethe-Salpeter equation and then study the nuclear dipole modes in
Ni-isotopes. The evolution of dipole strengths in both low-lying
region and normal GDR region in neutron rich Ni-isotopes is
discussed. In addition, the relationship of the neutron skin and
low-lying dipole resonance is investigated.

The paper is arranged as the follows. In Sec.II we briefly
introduce the formalism of the QRRPA in the response function
method. The calculated ground state properties, such as the
binding energy, the difference of neutron and proton rms radii as
well as the neutron and proton density distributions, in neutron
rich even-even Ni-isotopes are presented in Sec.III. In Sec.IV the
evolution of isovector dipole resonances in the low-lying region
as well as in the GDR region are shown and discussed. The
relationship of neutron skin and low-lying dipole resonance is
also investigated. Finally we give a brief summary in Sec.V.

\section{The quasiparticle relativistic random phase approximation}

It is important to take into account the pairing correlations in
the study of multipole collective excitations for open shell
nuclei. The QRRPA in the response function formalism is employed
at the present study, which has provided a convenient and useful
method to describe collective excited states of nuclear many-body
systems. For the detailed description of QRRPA based on ground
state of the RMF+BCS can be found in Ref.\cite{Cao032}. In this
section we briefly summarize the essential points of
Ref.\cite{Cao032}.

In the QRRPA calculation we first solve the Dirac equation and the
equations of meson fields in the coordinate space. For neutron
rich nuclei the contribution of the particle continuum to the
pairing correlations should be considered. In this work, a proper
treatment of the resonant continuum to pairing correlations has
been performed in the extended RMF+BCS approximation\cite{Cao031},
where the resonant continuum is solved with an asymptotic
scattering boundary condition. It has been shown that the simple
BCS approximation with a proper treatment of the resonant
continuum works well in the description of ground state properties
even for neutron rich nuclei\cite{Cao031,Gra01}. Then the positive
energy continuum states and negative energy states beyond  the
pairing active space are solved by expanding the nucleon spinors
in a complete set of bases, such as eigenfunctions of a spherical
harmonic oscillator potential.

When the pairing correlations are taken into account, the
elementary excitation in the pairing active space is a
two-quasiparticle excitation, rather than a particle-hole
excitation. Beyond the pairing active space the RRPA
configurations remain: a set of particle-hole pairs($ph$) and
negative energy particle-hole ($\overline{\alpha}$$h$) pairs,
where the particle $\overline{\alpha}$ is in the Dirac sea. The
unperturbed polarization operator in the QRRPA in the response
function formalism can be constructed in a similar way as done in
Ref.\cite{Ma97}:

\begin{widetext}
\vglue -0.50cm
\begin{eqnarray}
&&\Pi _0^R(P,Q;k,k^{\prime };E) \nonumber \\
&&=\frac{(4\pi)^{2}}{2L+1}\left\{\sum_{\alpha,\beta}(-1)^{j_{\alpha}+j_{\beta}}A_{\alpha\beta}\left[ \frac{%
\langle\overline{\phi}_\alpha\|P_L\|\phi
_\beta\rangle\langle\overline{\phi
}_\beta\|Q_L\|\phi_\alpha\rangle}{E-(E_\alpha+E_{\beta})+i\eta }-\frac{\langle\overline{\phi }_\beta\|P_L\|\phi _\alpha\rangle%
\langle\overline{\phi}_\alpha\|Q_L\|\phi _\beta\rangle}{E+(E_\alpha+E_{\beta})+i\eta }%
\right]\right. \nonumber \\
&&\left.~~~~~~~+\sum_{\alpha,\overline{\beta}}(-1)^{j_{\alpha}+j_{\overline{\beta}}}\upsilon_{\alpha}^{2}\left[
\frac{ \langle\overline{\phi}_\alpha\|P_L\|\phi _{\overline{\beta
}}\rangle\langle\overline{\phi}_{\overline{\beta}}\|Q_L\|\phi_h\rangle}{E-(E_\alpha+\lambda-\varepsilon_{\overline{\beta
}})+i\eta }-\frac{\langle\overline{\phi
}_{\overline{\beta}}\|P_L\|\phi _\alpha
\rangle\langle\overline{\phi}_\alpha\|Q_L\|\phi
_{\overline{\beta}}\rangle}{E+(E_\alpha+\lambda-\varepsilon_{\overline{\beta}})+i\eta
} \right]\right\}\ ~,  \label{eq4}
\end{eqnarray}
\end{widetext}

with
\begin{equation}
A_{\alpha\beta}=(u_\alpha\upsilon_\beta+(-1)^{L}\upsilon_{\alpha}u_\beta)^{2}(1+\delta_{\alpha
\beta})^{-1}~, \label{eq5}
\end{equation}
where $\upsilon^{2}_\alpha$ is the BCS occupation probability and
$u^{2}_\alpha=1-\upsilon^{2}_\alpha$.
$E_\alpha=\sqrt{(\varepsilon_\alpha-\lambda_n)^{2}+\Delta^{2}_n}$
is the quasiparticle energy, where $\lambda_n$ and $\Delta_n$ are
the neutron Fermi energy and pairing correlation gap,
respectively. In the BCS approximation, the $\phi_\alpha$ is the
eigenfunction of the single particle Hamiltonian with an
eigenvalue $\varepsilon_\alpha$. In Eq.(1), the terms in the first
square bracket represent the excitation with one quasiparticle in
fully or partial occupied states and one quasiparticle in the
partial occupied or unoccupied states. The terms in the second
square bracket describe the excitation between positive energy
fully or partial occupied states and negative energy states in the
Dirac sea. For unoccupied positive energy states outside the
pairing active space, their energies are
$E_\beta=\varepsilon_\beta-\lambda_n$, occupation probabilities
$\upsilon^{2}_\beta=0$ and $u^{2}_\beta=1$. For fully occupied
positive energy states, the quasiparticle energies and occupation
probabilities are set as $E_\alpha=\lambda_n-\varepsilon_\alpha$
and $\upsilon^{2}_\alpha=1$ in Eq.(1). The states in the Dirac sea
are not involved in the pairing correlations, therefore the
quantities $\upsilon_{\overline{\beta}}^{2}$ and
$u_{\overline{\beta}}^{2}$ are set to be $0$ and $1$. Once the
unperturbed polarization operator in the quasiparticle scheme is
built, the QRRPA response function can be obtained by solving the
Bethe-Salpeter equation as done in the RRPA\cite{Ma97}.

\section{ground state properties of neutron rich Ni-isotopes}

A reliable description on single particle energies and wave
functions and self-consistent calculation in the QRRPA is very
important. In this section we investigate the ground state
properties of neutron rich even-even Ni-isotopes in the extended
RMF+BCS approximation\cite{Cao031}, where the continuum is treated
by imposing a scattering boundary condition. Only those single
particle resonant states in the continuum are considered in the
pairing correlation, where
the width of the resonant state is not included.
The calculations are performed with parameter set NL3\cite{Lal97}.

\begin{figure}[hbtp]
\includegraphics[scale=0.5]{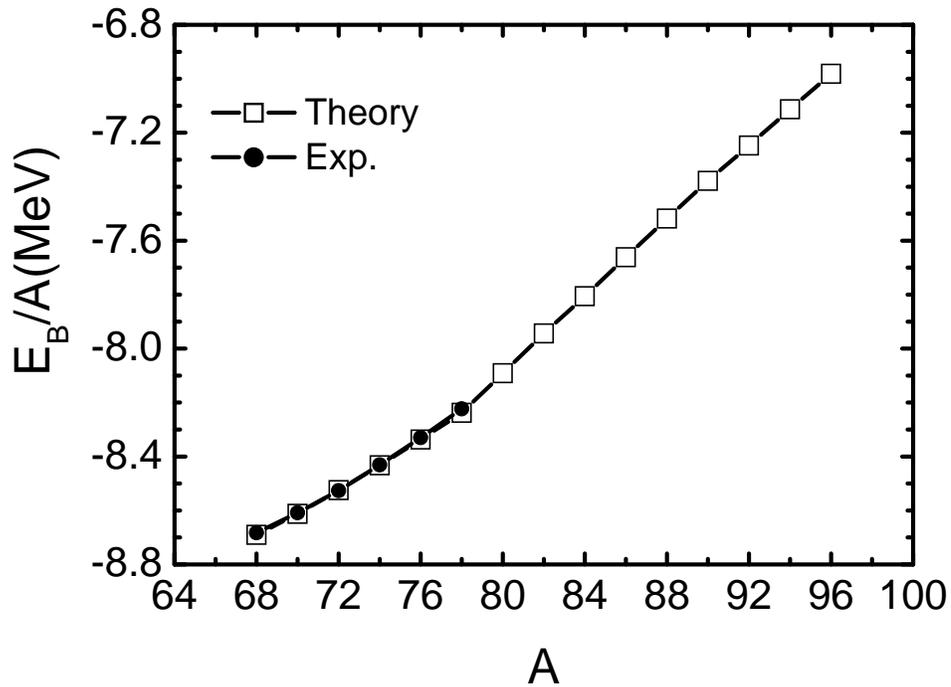}
\vglue -4.50cm\caption{The binding energy per nucleon of neutron
rich Ni-isotopes with mass from 68 to 96. The open squares
represent the results obtained by extended RMF+BCS approach with
effective interaction NL3. The experimental data are taken from
Ref.\cite{Audi95} denoted by solid circles.} \label{Fig.1}
\end{figure}

\begin{figure}[hbtp]
\includegraphics[scale=0.5]{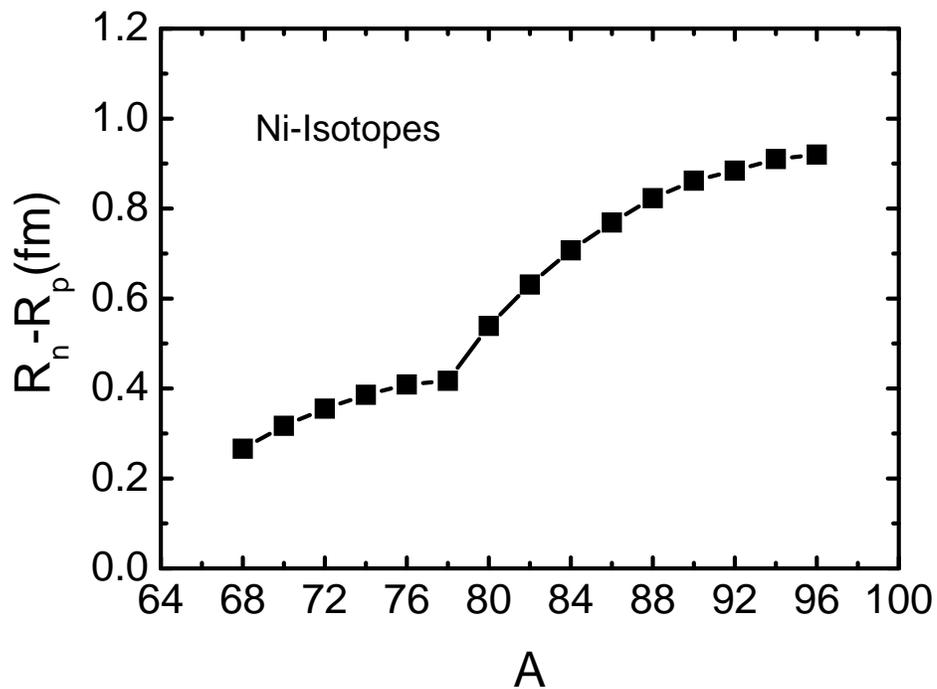}
\vglue -4.5cm\caption{The differences of neutron and proton rms
radii of neutron rich Ni-isotopes with mass from 68 to 96.}
\label{Fig.2}
\end{figure}

\begin{figure}[hbtp]
\includegraphics[scale=0.5]{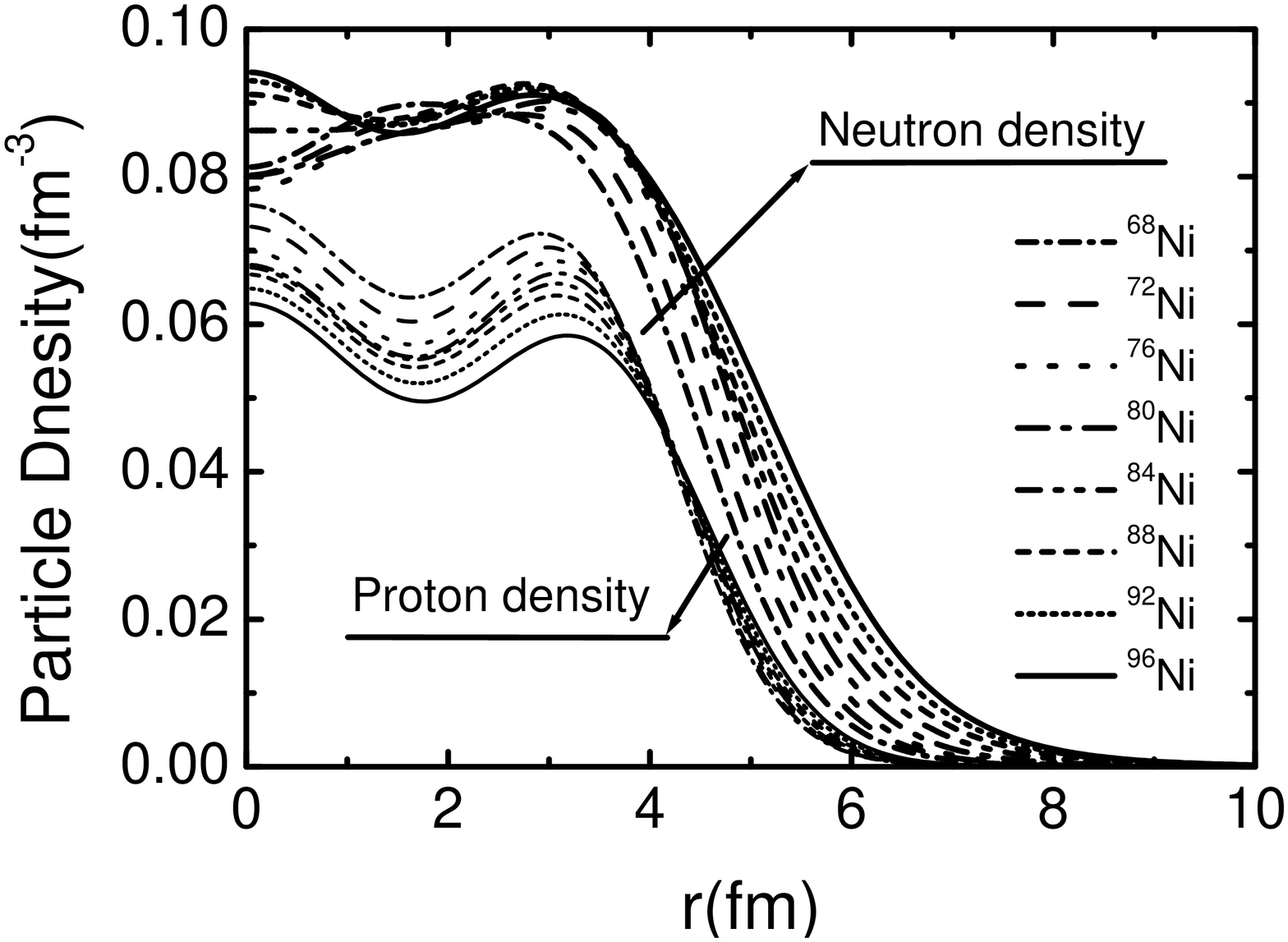}
\vglue -4.50cm\caption{The density distribution of neutron and
proton of neutron rich Ni-isotopes with mass from 68 to 96.}
\label{Fig.3}
\end{figure}

Ni isotopes have the proton number $Z=28$,
which is a closed shell. Therefore, the proton pairing gap is
taken to be zero. For the neutron case, we use a state-independent
pairing strength $G=C/A$, where the constant $C=20.5$ MeV is
adopted from Ref.\cite{Est01}. It could also provide a best fit to
reproduce experimental binding energies of known Ni-isotopes. In
practical calculations, we restrict the pairing space to about one
harmonic oscillator shell above and below the Fermi surface in the
extended RMF+BCS model. The levels include $1f_{5/2}$, $2p_{3/2}$,
$2p_{1/2}$,$1g_{9/2}$, $2d_{5/2}$, $3s_{1/2}$, $2d_{3/2}$,
$1g_{7/2}$, $1h_{11/2}$ and $2f_{7/2}$. Some of them, such as
$2d_{5/2},~ 2d_{3/2},~ 1g_{7/2}, ~ 2f_{7/2}$, and $1h_{11/2}$ are
single particle resonant states depending on the isotopes.

In Fig.1 we show the binding energy per nucleon for neutron rich
Ni-isotopes with mass from 68 to 96 and the available experimental
data\cite{Audi95}. One can find that the theoretical results
obtained in the extended RMF+BCS approach reproduce the
experimental data of binding energy perfectly. It has been
understood that the RMF plus simple BCS approximation can give a
good description on the ground state properties even for nuclei
far from the $\beta$ - stability line when the coupling of bound
states and particle continuum is treated
properly\cite{Gra01,Cao031}. The isotope shift, i.e. the
difference of the neutron and proton rms radius in neutron rich
Ni-isotopes is plotted as a function of the nuclear mass number in
Fig.2. It can been seen that the difference of the neutron and
proton rms radius becomes larger when the nucleus closes to the
neutron drip line. The kink appeared around $A=78$ is due to the
effect of shell structure where the neutron number equals to 50.

When one takes into account the effect of pairing correlations,
the nucleon density can be written as:

\begin{equation}
\rho(r)=\sum_\alpha\frac{(2j+1)}{4\pi}\upsilon_\alpha^{2}\phi_\alpha^{\dagger}(r)\phi_\alpha(r)
~, \label{eq6}
\end{equation}
which runs over all states weighted by the factor
$\upsilon_\alpha^{2}$. In Fig.3 we show the calculated particle
density distribution in neutron rich Ni-isotopes with the mass
number from 68 to 96. The neutron densities in those nuclei are
much larger than the proton densities. The neutron skins are
clearly formed, which contribute to the nuclear dipole modes at
low energies.

\section{evolution of dipole modes in neutron rich Ni-isotopes}

We now apply the QRRPA to investigate the evolution of isovector
giant dipole resonance(IVGDR) in neutron rich Ni-isotopes. The
isovector dipole excitation is an $L=1$ type electric
(spin-non-flip) $\Delta T=1$ and $\Delta S=0$ giant resonance with
spin and parity $J^\pi =1^{-}$. The electric isovector dipole
operator we used in present calculations is:
\begin{equation}
Q_{1m}=\frac{1}{2}e\sum_{i=1}^A(\tau_i^0-\langle\tau^0\rangle)r_iY_{1m}(\hat{r}_i),~
\label{eq7}
\end{equation}
where the average $\langle\tau^0\rangle$ is (N-Z)/A.

In the QRRPA calculation the particle-hole residual interactions
are taken from the same effective interaction NL3 as that in the
description of the ground states of neutron rich Ni-isotopes. The
fully occupied states and states in the pairing active space are
calculated self-consistently in the RMF+BCS approach in the
coordinate space. The unoccupied states outside of the pairing
active space are obtained by solving the Dirac equation in the
expansion on a set of harmonic oscillation basis. The response
functions of the nuclear system to the external operator are
calculated at the limit of zero momentum transfer. In order to
guarantee the conservation of the vector current, the space-like
parts of vector mesons in the QRRPA calculations are taken into
account.

\begin{figure}[hbtp]
\includegraphics[scale=0.5]{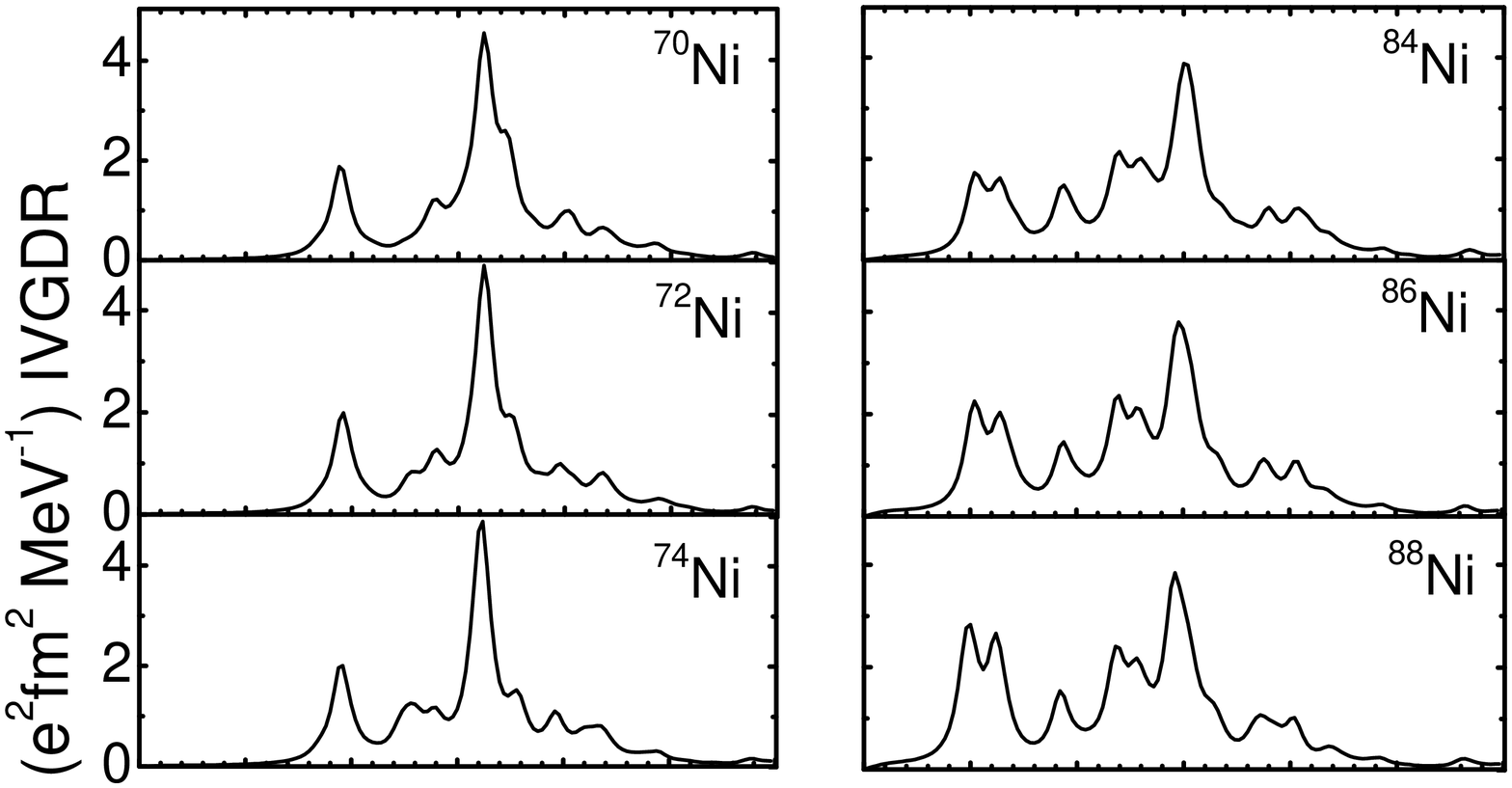}
\vglue -5.80cm
\includegraphics[scale=0.5]{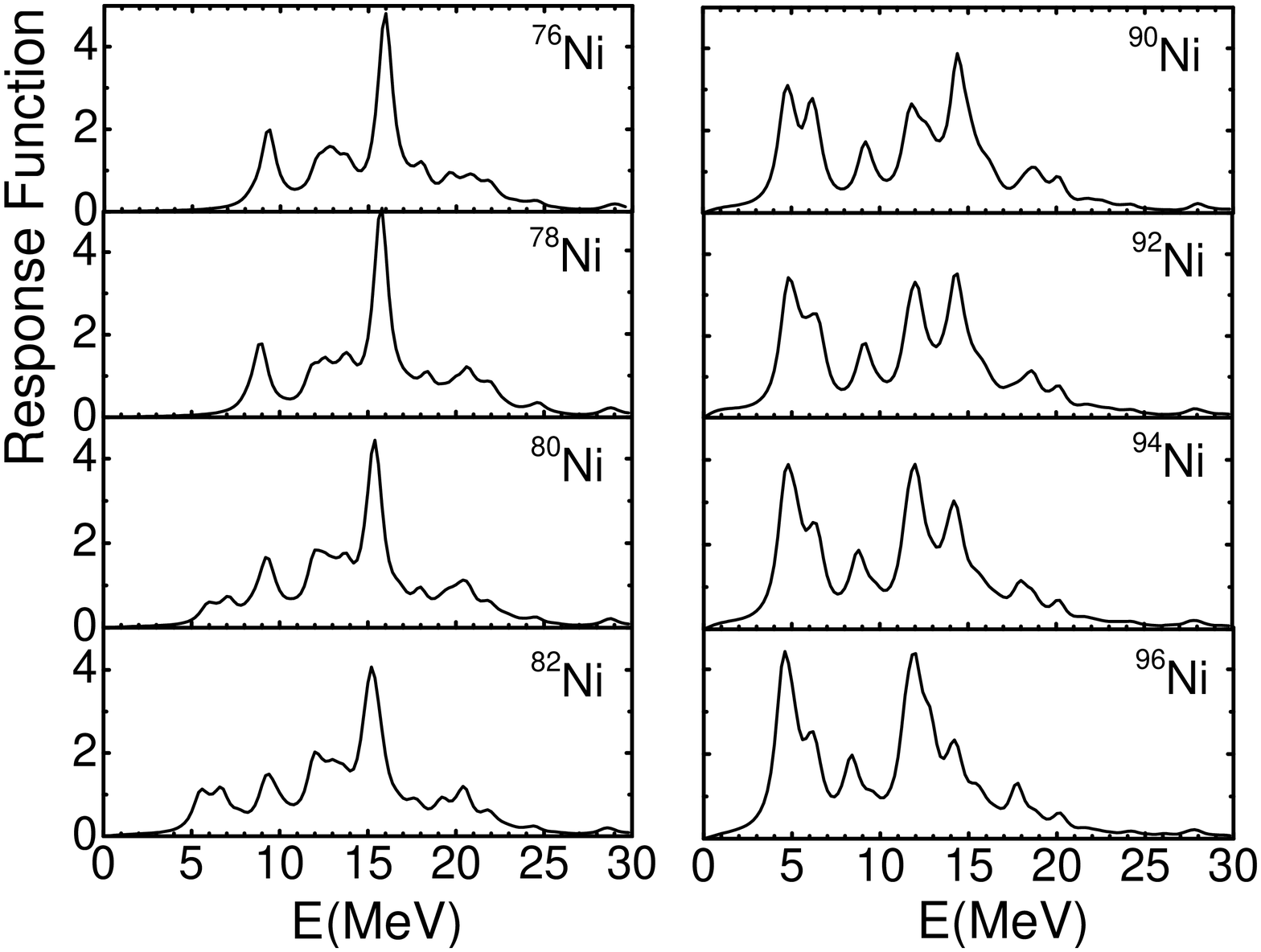}
\vglue -4.5cm \caption{The isovector dipole response functions in
neutron rich Ni-isotopes with the mass number from 70 to 96 in the
QRRPA approach.} \label{Fig.4}
\end{figure}

In Fig.4 we plot the response functions of isovector giant dipole
resonance in neutron rich Ni-isotopes calculated in the QRRPA
approach. It is found that the IVGDR strengths in those nuclei are
very fragmented, especially in nuclei near the neutron drip line.
In addition to the normal GDR strength around the energy at 16
MeV, the low-lying dipole strength appears at the excitation
energy below 10 MeV. The low-lying dipole strengths increase as
the neutron number increases. It is found that one pronounced peak
appears in the low energy region in nuclei with the mass number
less than 80. For those nuclei with the mass number larger than
80, one could observe more than one peak in the energy below 10
MeV. Moreover the strengths of those dipole excitations become
more and more stronger when the nucleus is approaching to the
neuron drip line.

\begin{figure}[hbtp]
\includegraphics[scale=0.5]{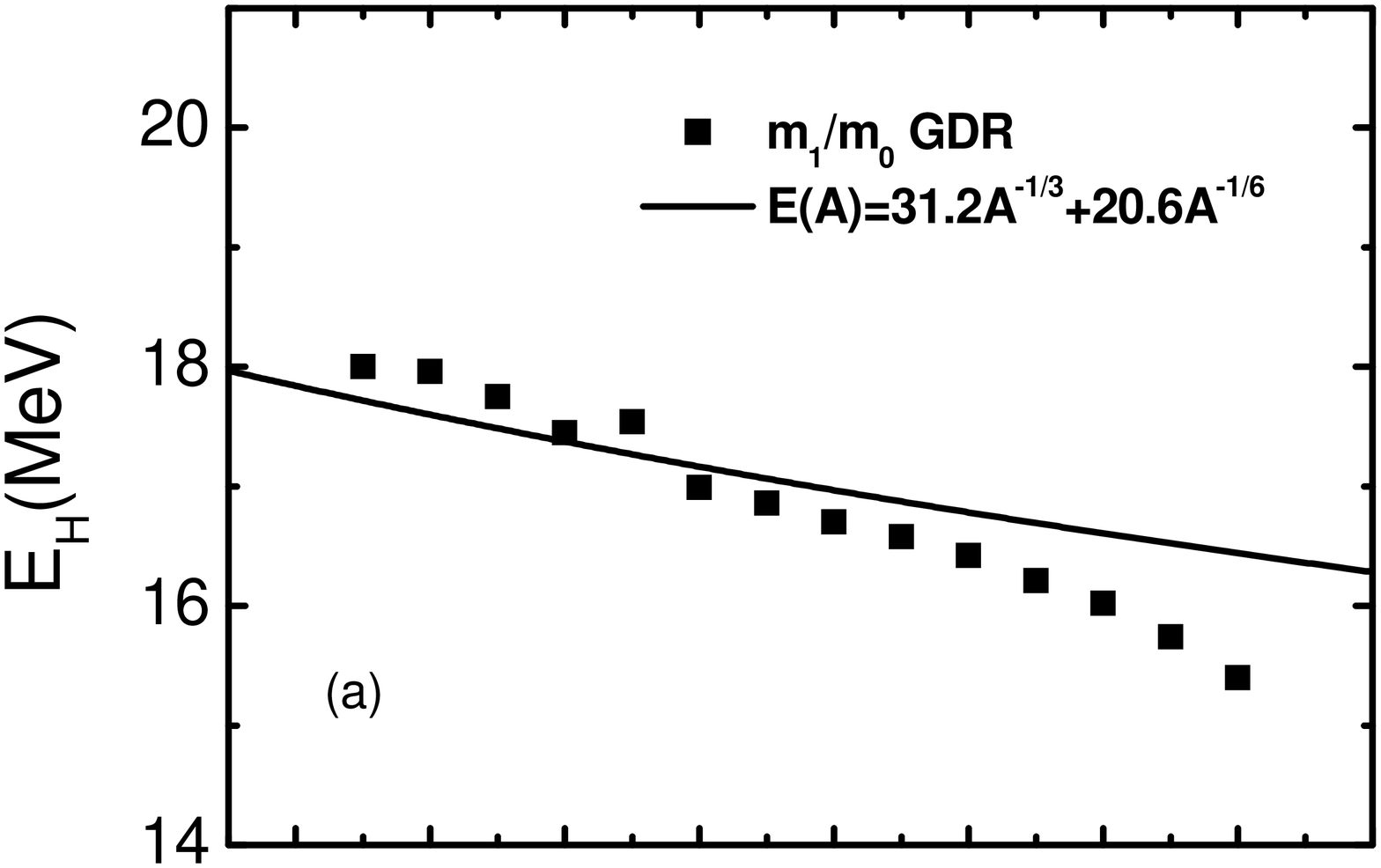}
\vglue -6.50cm
\includegraphics[scale=0.5]{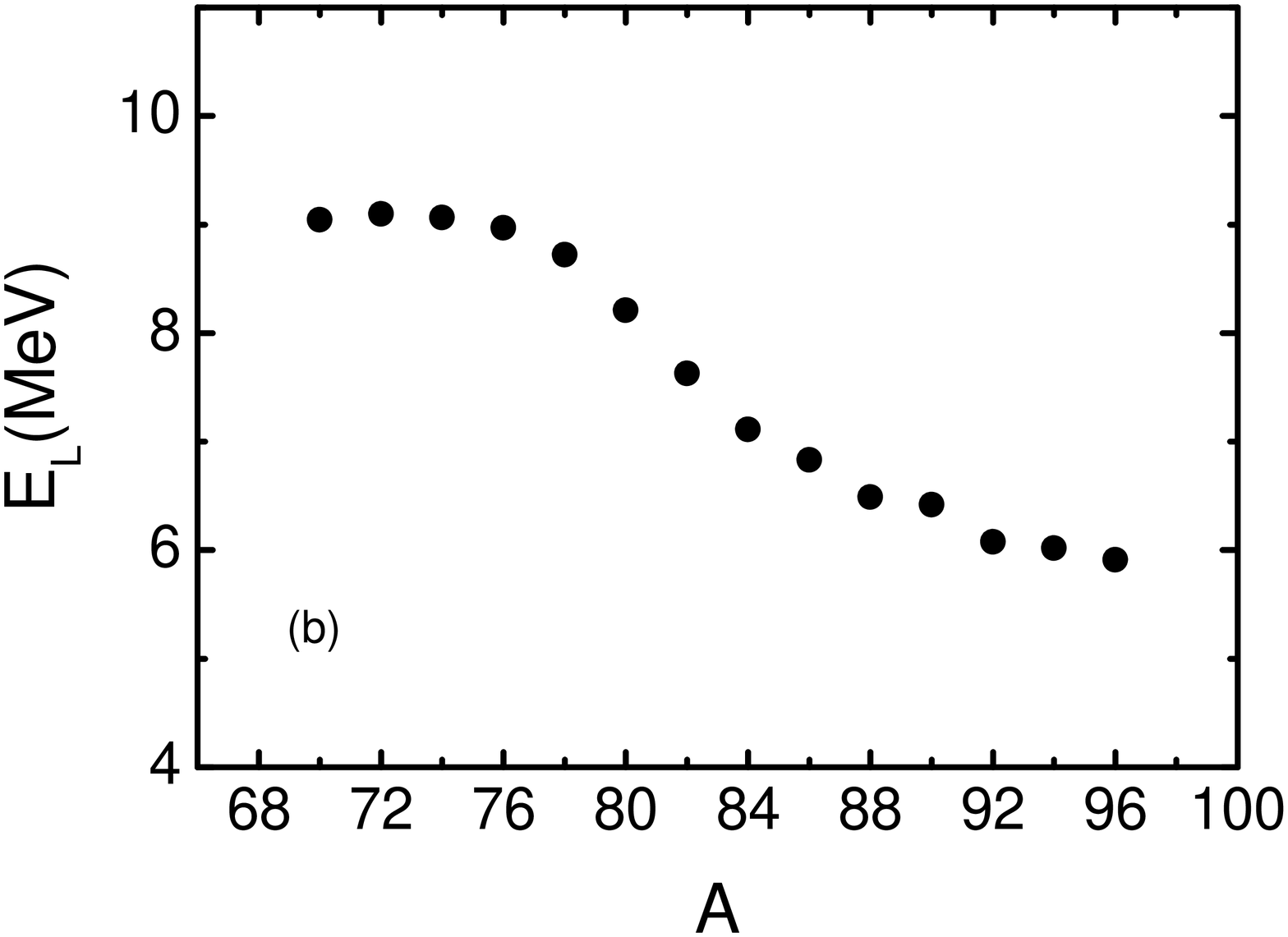}
\vglue -4.50cm \caption{The centroid energies of the IVGDR in the
normal GDR region (upper panel) and the low-lying region below 10
MeV (lower panel) as a function of the mass number in the
Ni-isotopes. The empirical expression of
$E=31.2A^{-1/3}+20.6A^{-1/6}$ for the normal GDR is plotted in the
upper panel, which is compared with the theoretical results. }
\label{Fig.5}
\end{figure}

In order to give clearer description on the evolution of those
dipole states in neutron rich Ni-isotopes, 
we calculate the
various moments of the IVGDR strengths in the QRRPA approach at a
given energy interval:
\begin{equation}
m_{k}=\int_{E_1}^{E_{2}}R^{L}(E^{'})E^{'k}dE^{'}~, \label{eq8}
\end{equation}
The RPA equation is solved till $E = 60$
MeV in the present calculations. The centroid energy of the
response function is defined as,
\begin{equation}
\overline{E}=m_{1}/m_{0}~, \label{eq9}
\end{equation}
we separate the energy interval to two regions: 0 MeV $< E_x <$ 10
MeV and 10 MeV $< E_x <$ 60 MeV. In the upper panel of Fig.5 the
calculated centroid energies of the normal GDR are plotted, which
are denoted by solid squares. It is compared with the empirical
expression of the GDR energy $E=31.2A^{-1/3}+20.6A^{-1/6}$, which
are displayed by a solid curve. One can see that the calculated
centroid energies are in good agreement with the empirical values
when the nuclei are not very far away from the $\beta$ stability
line. With the increase of the neutron excess, the calculated
centroid energies become smaller and deviate from the empirical
values. The calculated centroid energy of the IVGDR is 15.4 MeV
for $^{96}$Ni , which is about 1 MeV smaller than the
corresponding empirical value. The centroid energies of low-lying
dipole strengths are plotted by solid circles in the lower panel
of Fig.5. The centroid energies predicted in the QRRPA approach
decrease with the increase of the neutron excess.

\begin{figure}[hbtp]
\includegraphics[scale=0.5]{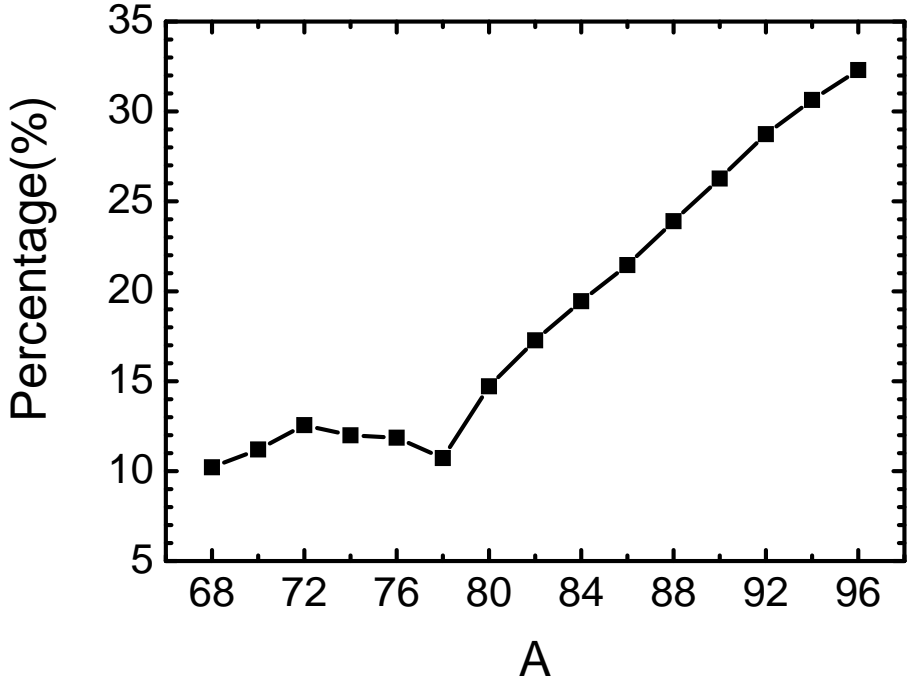}
\vglue -4.50cm\caption{The percentage of the calculated energy
weighted moments $m_1$ in the low-lying region ($E \leq 10$ MeV)
with respect to the corresponding Thomas-Reiche-Kuhn dipole sum
rule value as a function of the mass number of  Ni-isotopes. }
\label{Fig.6}
\end{figure}

Recently the experimental results show that the low-lying dipole
strength exhausts about 5\% of the Thomas-Reiche-Kuhn (TRK) dipole
sum rule for light neutron rich nuclei $^{18}$O, $^{20}$O and
$^{22}$O\cite{Aum96}. In the future experiments the low-lying
dipole excitation in more neutron rich medium-heavy nuclei will be
performed. Therefore we calculate the energy weighted moments
$m_1$ of the low-lying dipole strengths in the energy region 0 MeV
$< E_x <$ 10 MeV and the percentages to exhaust the corresponding
classical TRK dipole sum rule value. The values to exhaust the TRK
sum rule in various Ni-isotopes are shown in Fig.6. It is found
that the ratios of the energy weighted moments m$_{1}$ at the low
energy region increase on the whole as the increase of the neutron
excess. In present calculations, the contribution of low-lying
dipole strength in $^{70}$Ni exhausts 10.2\% of the classical TRK
dipole sum rule. Whereas for $^{96}$Ni which is close to the
neutron drip line, the low-lying dipole strength exhausts about
32.3\% of the TRK dipole sum rule. A similar results for isovector
dipole excitation in neutron rich nuclei has been obtained by D.
Vretenar \cite{Vre01}.

\begin{figure}[hbtp]
\includegraphics[scale=0.5]{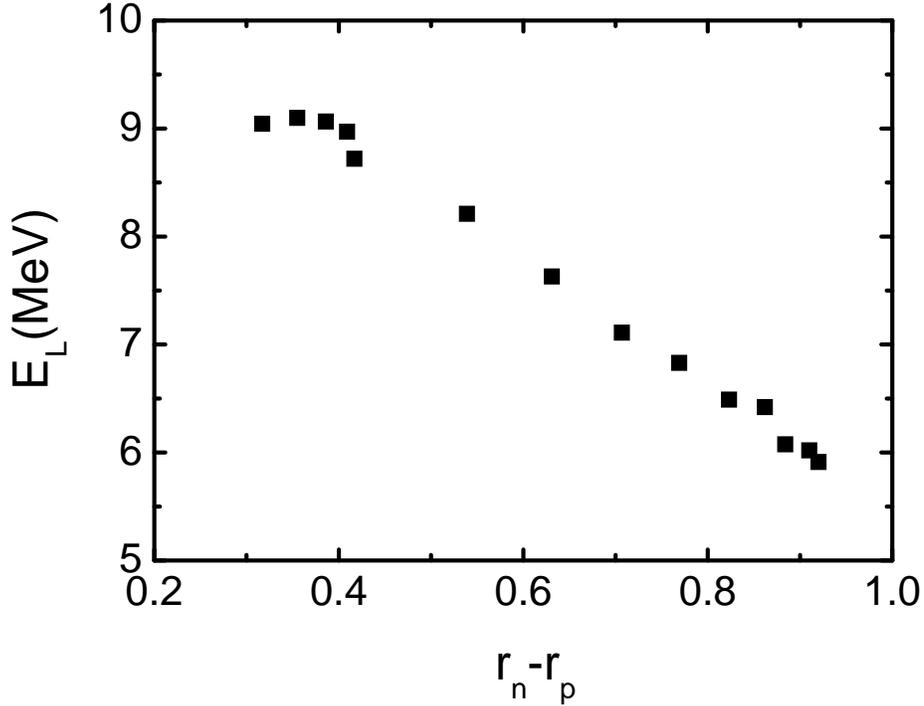}
\vglue -4.50cm\caption{The centroid energies of the isovector
dipole strengths in the low-lying region below 10 MeV versus the
differences of the neutron and proton rms radii in Ni-isotopes.}
\label{Fig.7}
\end{figure}

Differing from the normal GDR response, the low-lying resonance
can be interpreted as the vibration of the excess neutrons against
the core formed with equal number of protons and neutrons out of
phase\cite{Vre01}. The recent experimental results show that one
can extract the neutron skin by measuring the dipole excitation of
nuclei\cite{Kra91,Kra99}.  The calculated centroid energies of
low-lying dipole strengths as a function of the differences of the
neutron and proton rms radii in Ni-isotopes is plotted in Fig.7.
We notice that the evolution of centroid energies in the low
energy region has a strong dependence on the thickness of neutron
skin. This relationships was also noted and discussed in
Ref.\cite{Tso03}, where the low-lying dipole strength distribution
in neutron rich nuclei was calculated in a quasiparticle phonon
model. They found that the transition strengths and energy
locations of low-lying dipole resonances are closely correlated
with the neutron skin. Therefore, the experimental measurements on
the low-lying dipole excitation in neutron-rich nuclei would
provide information on the neutron skin.

\section{summary}

In the present work we have studied the evolution of the low-lying
dipole resonances in neutron rich even-even Ni-isotopes in the
framework of the QRRPA with an effective Lagrangian parameter set
NL3. The pairing correlations are taken into account within the
BCS approximation with a constant pairing gap. The distributions
of dipole excitation in neutron rich nuclei are more fragmented
and the soft mode of the dipole response is found at the
low-energy region. The energy weighted moments $m_{1}$ at the
low-lying region calculated in the QRRPA approach increase as the
increase of the neutron excess. A considerably large percentage of
the low-lying $m_{1}$ to exhaust the TRK sum rule, is found when
the nuclei approach to the neutron drip line. The calculated
centroid energies of normal GDR strengths is in good agreement
with the empirical values when the nucleus is not very far away
from the $\beta$ stability line. Whereas a deviation from the
empirical value can be observed in nuclei close to the drip line.
The centroid energies of low-lying dipole strengths decrease with
the increase of the neutron excess. An interesting phenomenon is
that the evolution of centroid energies at the low energy region
has a strong dependence on the thickness of the neutron skin in
neutron rich even-even Ni-isotopes. Therefore, the experimental
measurement of the nuclear low-lying dipole strength distributions
in neutron-rich nuclei is suggested, which will provide the
information of neutron skin.

\begin{acknowledgments}
One of the authors (LGC) wishes to thank Prof. Zhang Zong-ye and
Prof. Yu You-wen for many stimulating discussions. This work is
supported by the National Natural Science Foundation of China
under Grant Nos 10305014, 90103020, and 10275094, and Major State
Basic Research Development Programme in China under Contract No
G2000077400.
\end{acknowledgments}

\end{document}